%% file: SciPost_LaTeX_Template.tex
\documentclass[submission, Proceedings]{SciPost}

\hypersetup{
    colorlinks,
    linkcolor={red!50!black},
    citecolor={blue!50!black},
    urlcolor={blue!80!black}
}

\usepackage[bitstream-charter]{mathdesign}
\usepackage{xspace}
\usepackage{graphicx}
\usepackage{placeins}
\usepackage{subcaption}
\graphicspath{ {./figures/} }

\DeclareSymbolFont{usualmathcal}{OMS}{cmsy}{m}{n}
\DeclareSymbolFontAlphabet{\mathcal}{usualmathcal}

\begin{document}

\begin{center}{\Large \textbf{
Dark Sector searches with jets\\
}}\end{center}

\begin{center}
  J. von Ahnen\textsuperscript{1$\star$}, on behalf of the ATLAS and CMS Collaborations
\end{center}

\begin{center}
{\bf 1} DESY, Hamburg, Germany
\\
* janik.von.ahnen@cern.ch
\end{center}

\begin{center}
\today
\end{center}


\definecolor{palegray}{gray}{0.95}
\begin{center}
\colorbox{palegray}{
  \begin{tabular}{rr}
  \begin{minipage}{0.1\textwidth}
    \includegraphics[width=30mm]{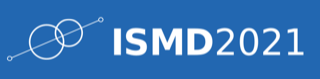}
  \end{minipage}
  &
  \begin{minipage}{0.75\textwidth}
    \begin{center}
    {\it 50th International Symposium on Multiparticle Dynamics}\\ {\it (ISMD2021)}\\
    {\it 12-16 July 2021} \\
    \doi{10.21468/SciPostPhysProc.?}\\
    \end{center}
  \end{minipage}
\end{tabular}
}
\end{center}

\section*{Abstract}
{\bf
The presence of a non-baryonic Dark Matter (DM) component in the Universe is
inferred from the observation of its gravitational interaction. The ATLAS and
CMS experiments located at the LHC have developed a broad search program for DM
candidates, including resonance searches for the mediator and searches with
large missing transverse momentum. Additionally, searches have been conducted in
models where the Higgs Sector and the Dark Sector are connected leading for
example to invisible Higgs boson decays. The results of recent searches on 13
TeV $pp$ data, their interplay and interpretation are presented.
}

\input{main_body}

\bibliography{my_bib}

\nolinenumbers

\end{document}

%% file: main_body.tex
\newcommand*{\MET}{\ensuremath{E_{\text{T}}^{\text{miss}}}\xspace}
\newcommand*{\pT}{\ensuremath{p_{\text{T}}}\xspace}
\newcommand*{\pTrecoil}{\ensuremath{p_{\text{T}}^{\text{recoil}}}\xspace}
\newcommand*{\BRinv}{\ensuremath{\mathcal{BR}_{\text{inv}}}\xspace}
\newcommand*{\GeV}{\ensuremath{\text{Ge\kern -0.1em V}}}

\section{Introduction}
\label{sec:intro}

Several astrophysical observations exist that can not be explained with the particle contents of the standard model (SM) of particle physics~\cite{DMreview2021}.
One approach to explain these observations is the introduction of non-baryonic Dark Matter (DM), which could explain the phenomenons with its gravitational impact.
This DM does not interact via the strong or the electromagnetic force.

There are three different categories of experiments that try to detect this DM.
One of these is the category of indirect detection experiments in which cosmic particles are investigated to look for hints of DM annihilation.
The second category is direct detection.
Here one looks for DM by measuring the recoil of SM particles coming from potential scattering with DM.
The last type of search looks for the production of DM using particle colliders.

The LHC is a proton-proton collider located at CERN in Geneva.
With its high centre-of-mass energy of 13 TeV, it could produce DM in its collisions.
ATLAS~\cite{2008ATLASdetector} and CMS~\cite{2008CMSdetector} are two general-purpose detectors at the LHC~\cite{2008LHCmachine} and have a broad program of searches for DM.
In high energy collisions of protons, the presence of jets is very likely.
Therefore, it is crucial to explore and understand final states with jets.

To interpret and guide analyses, models that extend the SM can be a good tool.
Simplified models are minimal extensions of the SM that generally have a rich phenomenology and small number of free parameters.

\section{\texorpdfstring{\MET}{ETmiss}+X searches}

Detectors like ATLAS or CMS can not directly detect DM.
Therefore, to identify events with DM, other SM particles have to be present.
These events lead to event topologies in which there is an imbalance in the transverse momentum (\MET).
Analyses that target these events are called \MET+X searches.
The \MET+jet search is one of these~\cite{ATLASmonojet,CMSmonojet}.

\subsection{\MET+jet strategy}

The signal signature has large \MET ($\gtrapprox 200 \GeV$) recoiling against a jet with high transverse momentum.
No leptons are expected in the event.

The major backgrounds for this signature are Z+jets, W+jets and $t\bar{t}$-production.
To minimize the uncertainties on the prediction of these backgrounds, control regions (CR) are constructed.
In these CRs the lepton requirements are changed in order to increase the contribution from backgrounds and discard the majority of signal events.

\subsection{\texorpdfstring{\MET}{ETmiss}+jet results}

\begin{figure}[h!]
  \centering
  \begin{subfigure}{.49\textwidth}
    \centering
    \includegraphics[width=\textwidth]{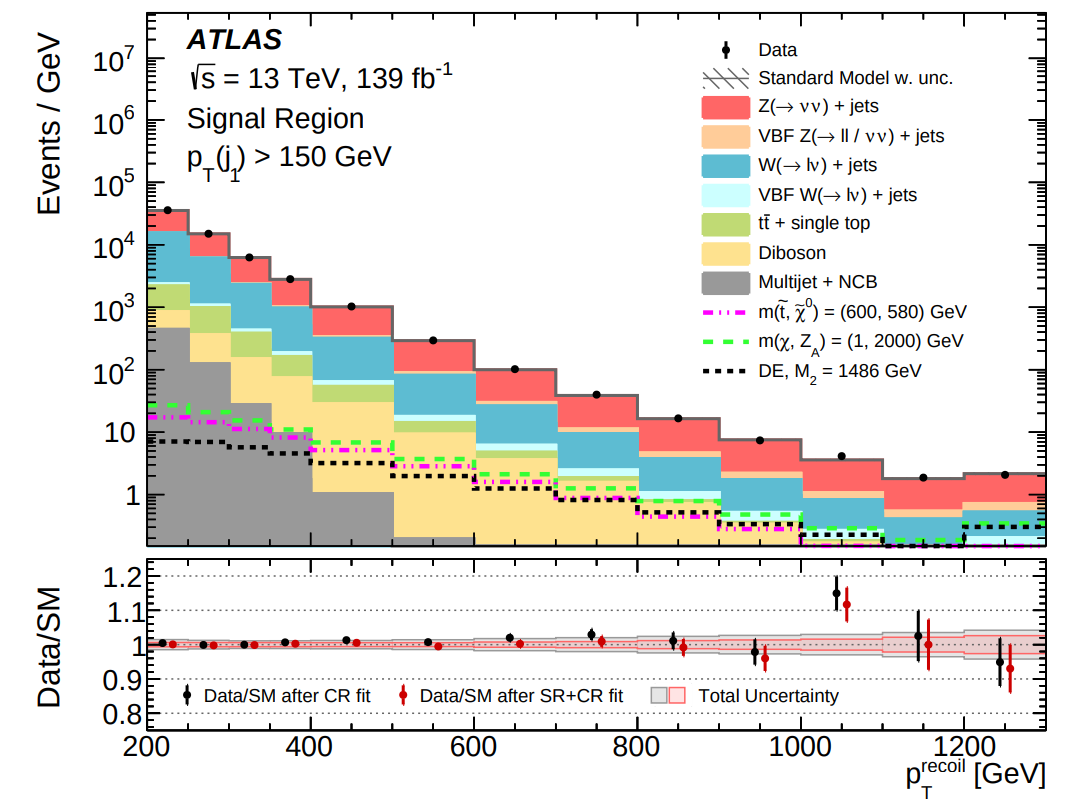}
    \caption{}
    \label{fig:monojetPtRecoil}
  \end{subfigure}%
  \hfill
  \begin{subfigure}{.49\textwidth}
    \centering
    \includegraphics[width=\textwidth]{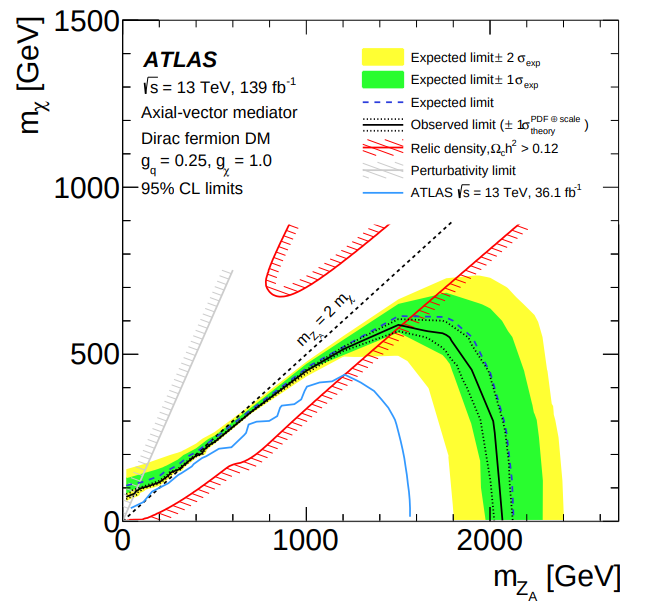}
    \caption{}
    \label{fig:monojetContour}
  \end{subfigure}
  \caption{(\subref{fig:monojetPtRecoil}) Measured distributions of \pTrecoil compared with the SM prediction in the SR. (\subref{fig:monojetContour}) 95\% CL exclusion contours in the $m_{Z_{A}}-m_{\chi}$ parameter plane for the axial-vector mediator model~\cite{ATLASmonojet}.}
  \label{fig:monojetresults}
\end{figure}

Figure~\ref{fig:monojetPtRecoil} shows the post fit distribution of the total transverse momentum of all visible objects (\pTrecoil).
This analysis can be interpreted in the context of different beyond the SM theories, e.g. invisible Higgs boson decays, large extra spatial dimensions or supersymmetric particles.
Figure~\ref{fig:monojetContour} shows the exclusion contour on a simplified model with an Axial-vector mediator.

\section{Resonance searches}

A new mediator that couples both to SM and DM particles, and that can be produced in collisions at the LHC would also have a resonant signature.
This would lead to an increased number of events where the collision energy is close to the mediator mass.

\subsection{Dijet resonance search strategy}

The dijet resonance search analysis looks for the enhanced number of events in the dijet mass spectrum~\cite{ATLASdijetresonance,CMSdijetresonance}.
In the SM this is a smoothly falling spectrum and can therefore be described by an analytical function.
A new resonance would create a bump on this spectrum.

\subsection{Results}

\begin{figure}[h!]
	\centering
	\includegraphics[width=0.85\linewidth]{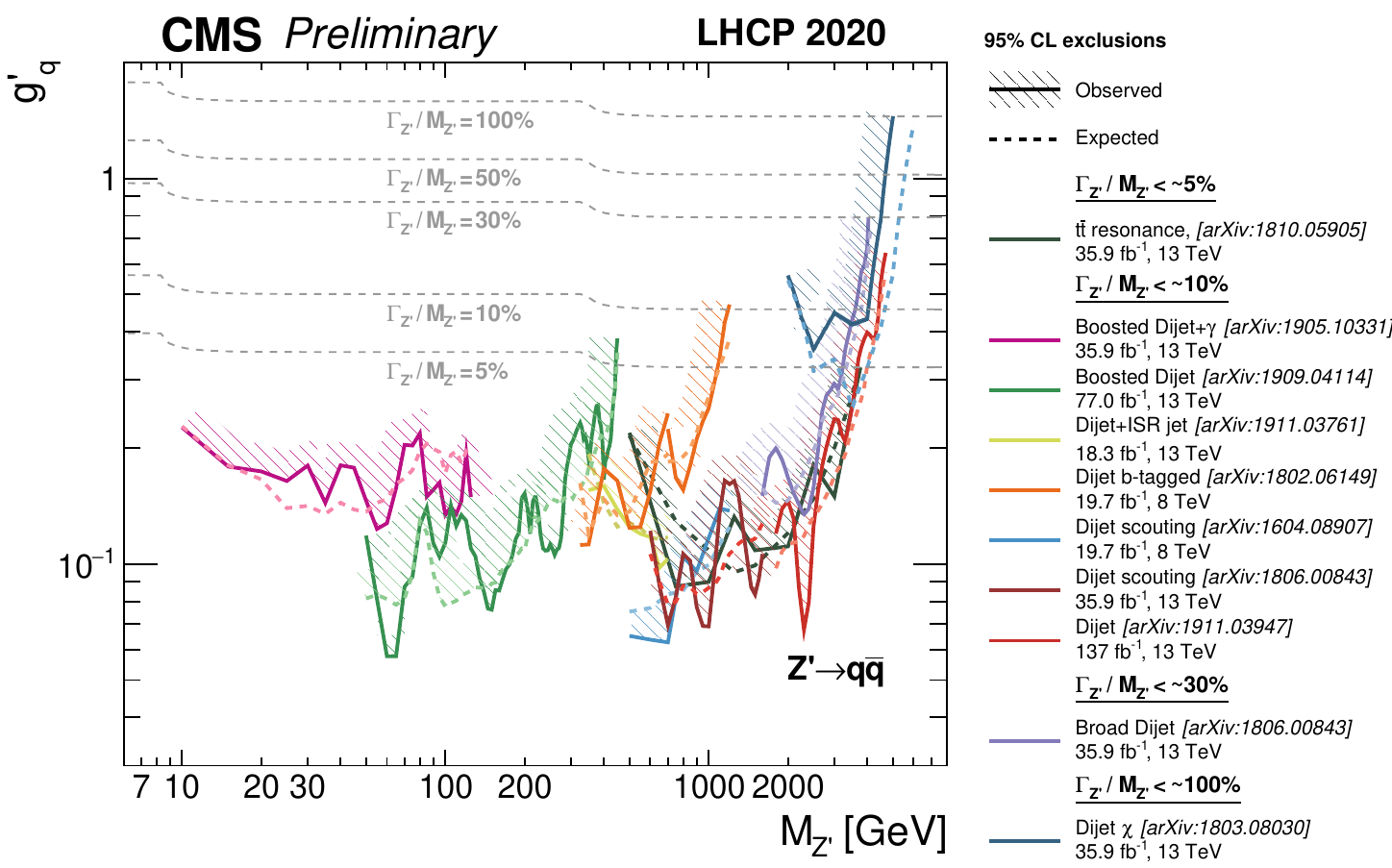}
	\caption{Limits on the universal coupling $g'_{q}$ between a leptophobic $Z'$ boson and quarks~\cite{dijetResonanceModels} from various CMS dijet analyses~\cite{CMSDMsummaryplots}.}
	\label{fig:dijetResonanceSummary}
\end{figure}

Because of the high event rate for dijet events in the low dijet mass region ($\lessapprox 1.5\,\mathrm{TeV}$), not all collisions can be recorded.
This leads to a limited sensitivity for the dijet resonance analysis relying on two separated jets.
By for example requiring additional objects in the events (e.g. photons) also the lower dijet mass region can be explored.
Figure~\ref{fig:dijetResonanceSummary} shows a summary plot for dijet resonance searches produced by the CMS collaboration.

\begin{figure}[h!]
  \centering
  \begin{subfigure}{.49\textwidth}
    \centering
    \includegraphics[width=\textwidth]{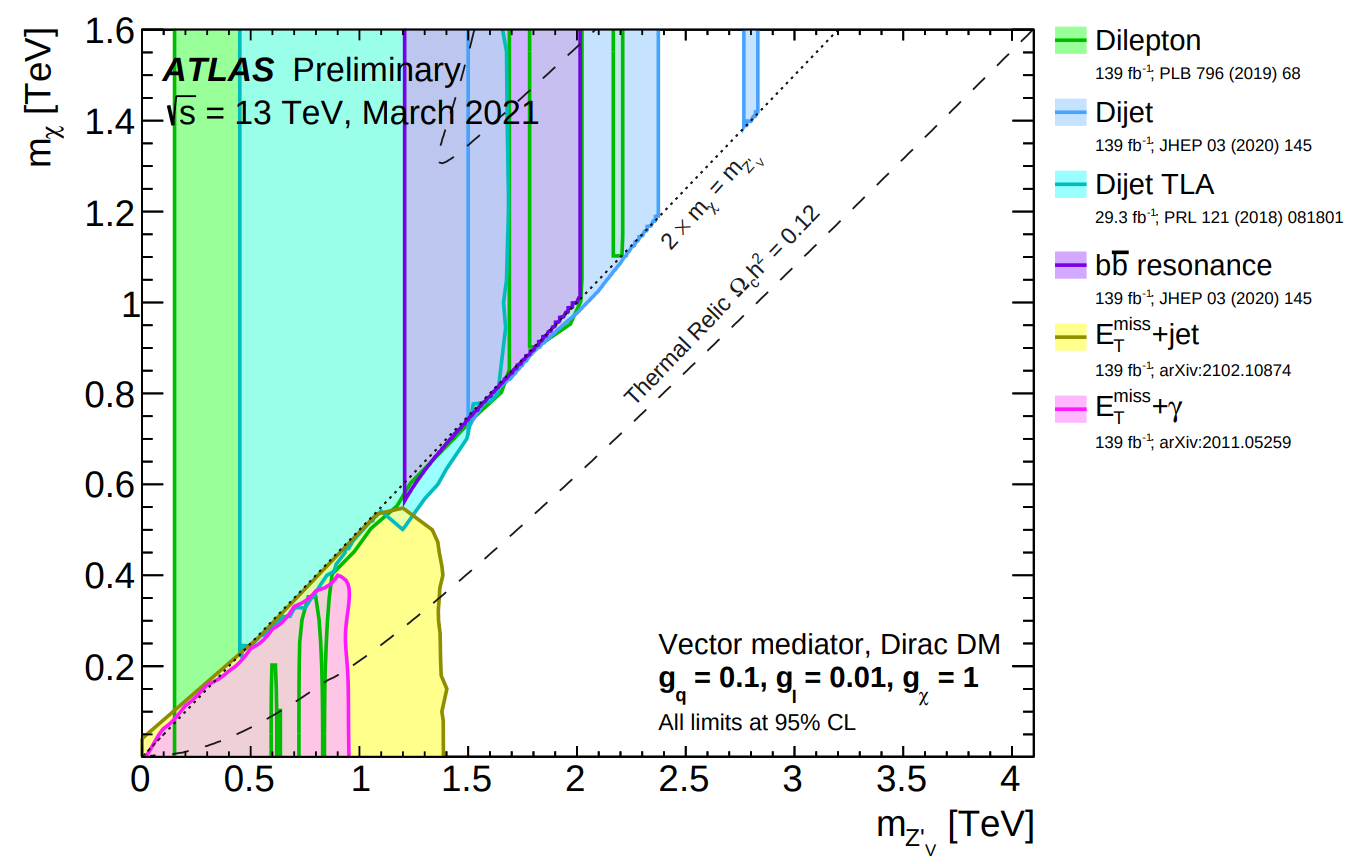}
    \caption{}
    \label{fig:dijetAndMonoX}
  \end{subfigure}%
  \hfill
  \begin{subfigure}{.49\textwidth}
    \centering
    \includegraphics[width=\textwidth]{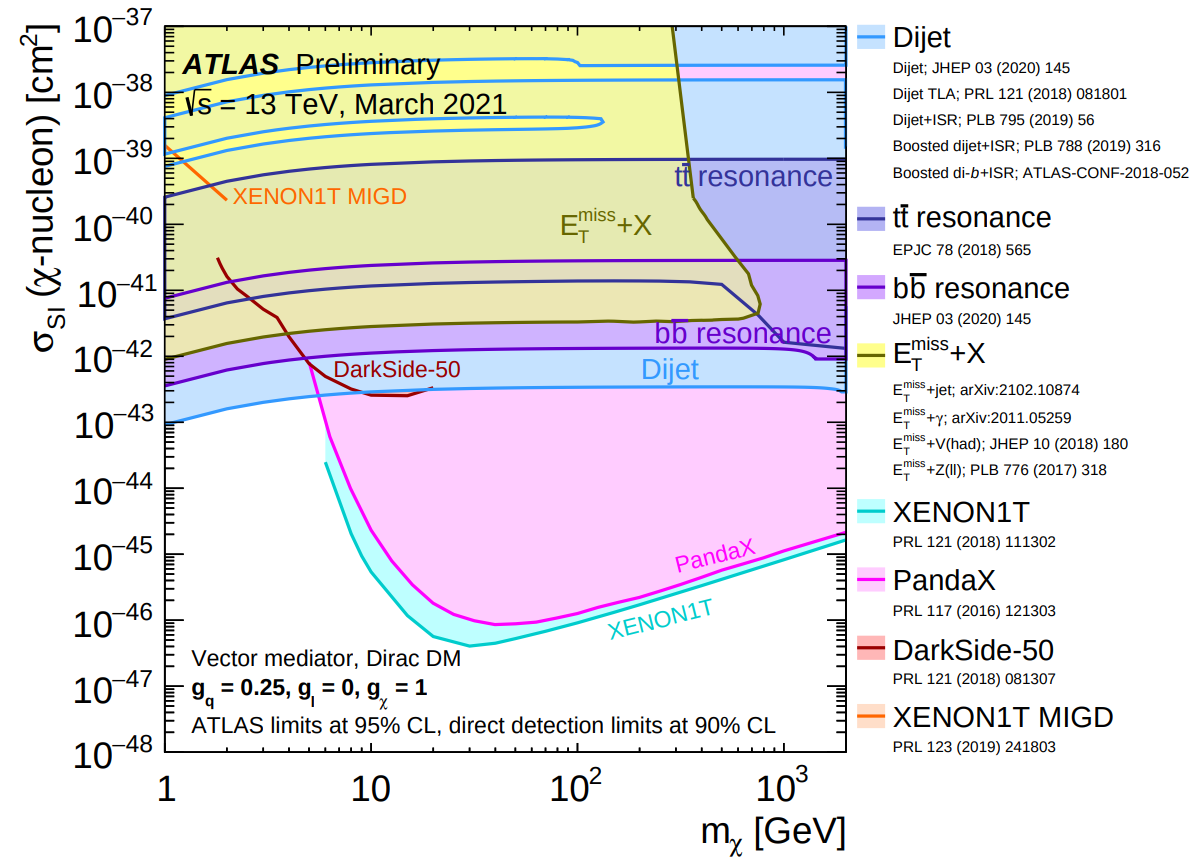}
    \caption{}
    \label{fig:DMjetAndDD}
  \end{subfigure}
  \caption{(\subref{fig:dijetAndMonoX}) 95\% CL exclusion contours for dijet, dilepton and \MET+X searches for leptophilic vector mediator simplified models. (\subref{fig:DMjetAndDD}) Comparison of spin-independent WIMP-nucleon scattering cross-section for various DM searches performed by ATLAS and limits from direct detection experiments~\cite{ATLASDMsummaryplots}.}
\end{figure}

The dijet analyses and the \MET+X searches provide a complementary way of exploring certain simplified models.
Figure~\ref{fig:dijetAndMonoX} shows the 95\% exclusion contours for a simplified model with a spin-1 mediator.
The exclusion contours depend on the choice of parameters.

The limits on the mediator mass can also be translated into DM-nucleon scattering cross-section limits which allows the results to be compared with those from direct detection experiments as shown in figure~\ref{fig:DMjetAndDD}.
Again the exclusion contours depend on the assumptions and choice of parameters.

\section{Search for invisible Higgs boson decays}

The Higgs boson couples to all known particles with mass and could therefore also couple to DM.
Models in which the Higgs boson connects the SM particles with the Dark Sector are called Higgs-portal models.
In this case, the branching ratio of the Higgs boson to invisible particles (\BRinv) would be larger than predicted by the SM.
In general, analyses focus on one of the four Higgs boson production modes (gluon-gluon fusion, vector-boson fusion (VBF), $t\bar{t}$-fusion, Higgsstrahlung) and measure \BRinv.
The analyses focusing on the VBF production are the most sensitive~\cite{ATLASVBFHinv,CMSVBFHinv}.

\subsection{VBF-H(inv) strategy}

The signal signature consists of two jets with a large separation in pseudorapidity ($\eta \gtrapprox 3.8$) and a high invariant dijet mass ($m_{jj} \gtrapprox 800\,\mathrm{GeV}$).
No isolated leptons are expected and the invisible Higgs boson decay results in large \MET ($\gtrapprox 200\,\GeV$).
The major backgrounds are W+jets and Z+jets production.
These are constraints by CRs, which are constructed by inverting the lepton veto of the SR.

\subsection{Results}

\begin{figure}[h!]
  \centering
  \begin{subfigure}{.45\textwidth}
    \centering
    \includegraphics[width=\textwidth]{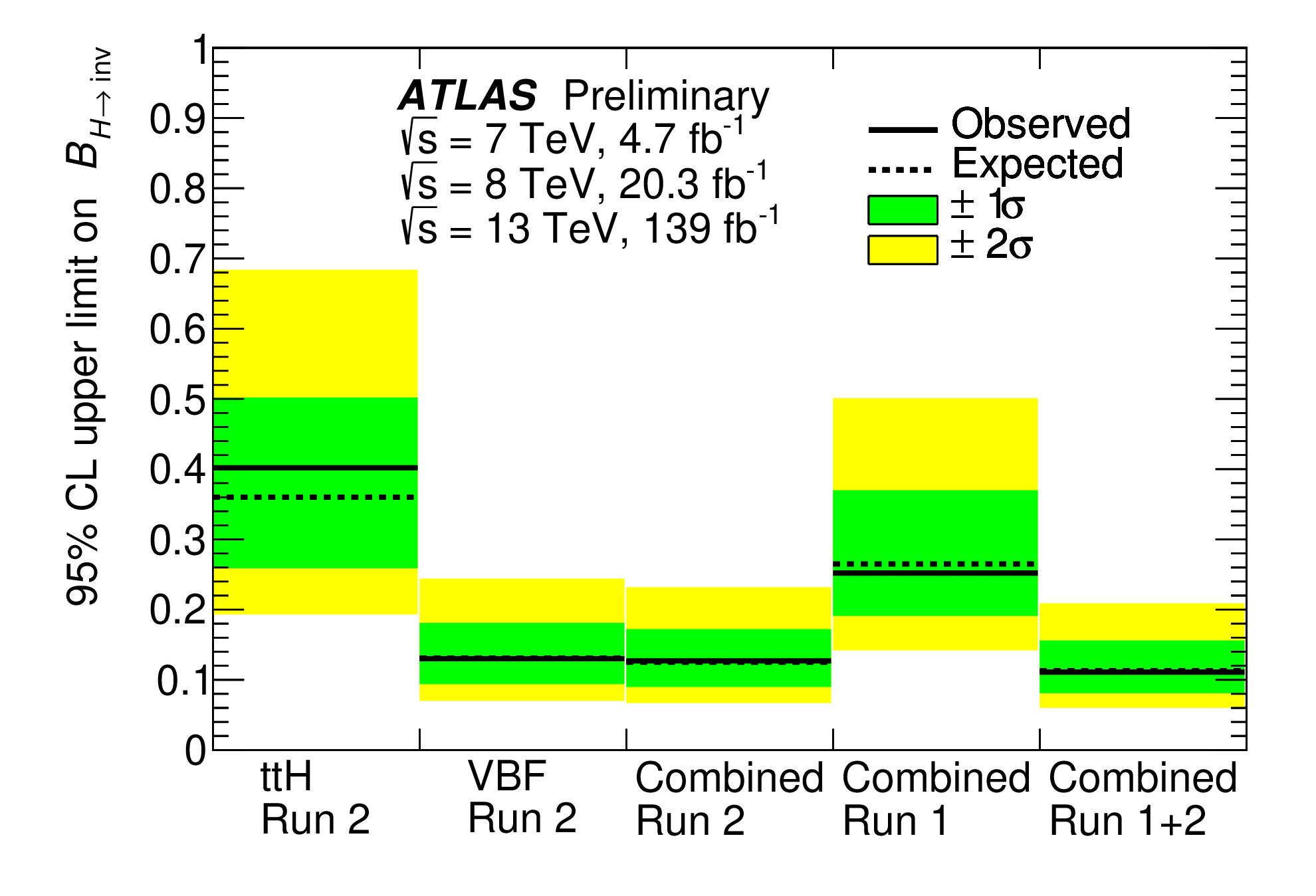}
    \caption{}
    \label{fig:HinvCombination}
  \end{subfigure}%
  \hfill
  \begin{subfigure}{.45\textwidth}
    \centering
    \includegraphics[width=\textwidth]{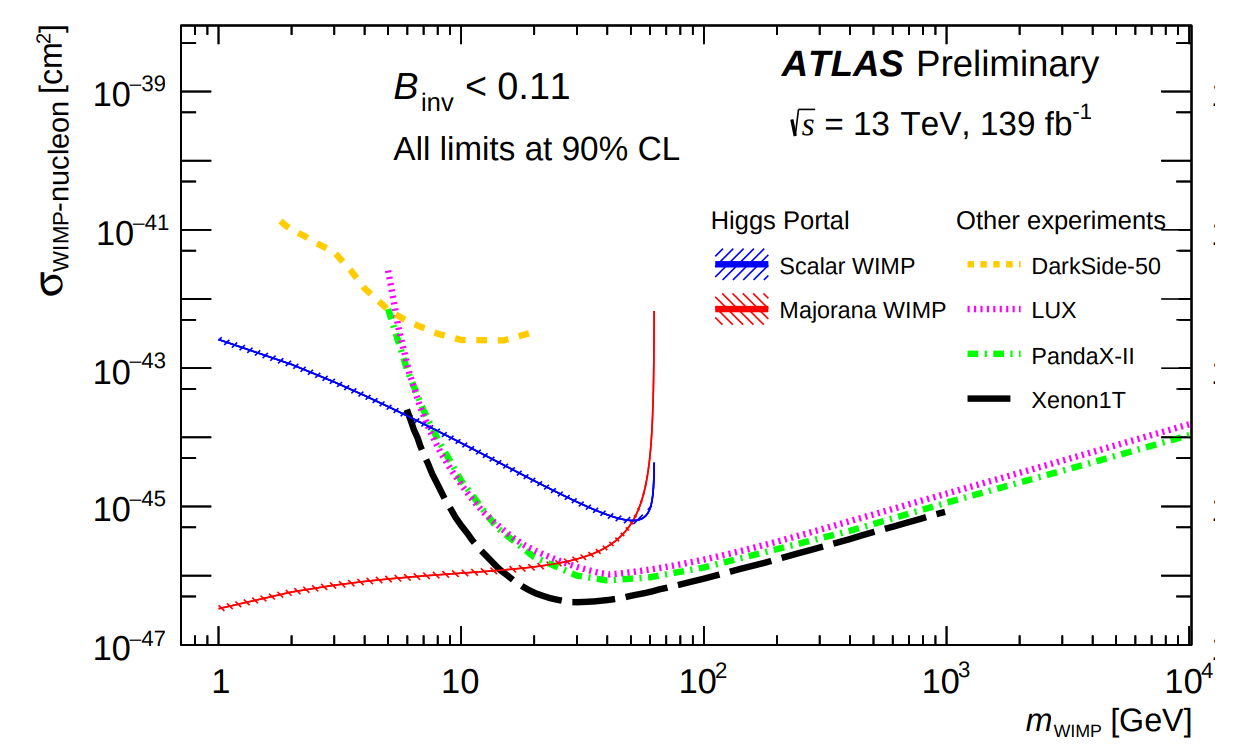}
    \caption{}
    \label{fig:HinvAndDD}
  \end{subfigure}
  \caption{(\subref{fig:HinvCombination}) Preliminary combination of Run 1 and Run 2 ATLAS results for invisible Higgs boson decays~\cite{ATLASHinvPrelimComb}. (\subref{fig:HinvAndDD}) Comparison of upper limits on the WIMP-nucleon scattering cross-section for direct detection experiments and a Higgs portal interpretation~\cite{ATLASVBFHinv}.}
\end{figure}

No excess from the SM is observed, and upper limits on \BRinv are set.
The best upper limit ($\BRinv > 11\%$) comes from the preliminary combination of Run 1 and Run 2 ATLAS results (see figure~\ref{fig:HinvCombination})~\cite{ATLASHinvPrelimComb}.
With certain assumptions, the upper limit on \BRinv can be translated to a WIMP-nucleon scattering cross-section and then compared to the limits achieved by direct detection experiments (see figure~\ref{fig:HinvAndDD}).

\section{Models with extended Higgs sector}

Several extensions of the SM involve changes to the Higgs sector.
The Two-Higgs-Doublet model (2HDM) extends the SM with a second Higgs doublet~\cite{2HDMpaper}.
One of the motivations for the 2HDM is that it could solve the problem of the observed baryon asymmetry in the universe.
The analysis focusing on DM produced in association with a SM Higgs boson decaying to $b$-quarks is sensitive to these models~\cite{ATLASmonoH,CMSmonoH}.

\subsection{\texorpdfstring{Mono-H($bb$)+\MET}{Mono-H(bb)+MET} strategy}

The signal signature is two $b$-jets with an invariant mass close to the SM Higgs boson mass ($\approx 125\,\mathrm{GeV}$).
Furthermore, this di-$b$-jet system is recoiling against large \MET ($\gtrapprox 150\,\mathrm{GeV}$) and no leptons are present in the event.

The major backgrounds are Z+jets, W+jets and $t\bar{t}$-production.
By inverting the lepton requirements of the SR, two CRs are defined and included in the fit to constrain the normalisation of the major backgrounds.

\subsection{\texorpdfstring{Mono-H($bb$)+\MET}{Mono-H(bb)+MET} results}

\begin{figure}[h!]
	\centering
	\includegraphics[width=0.49\linewidth]{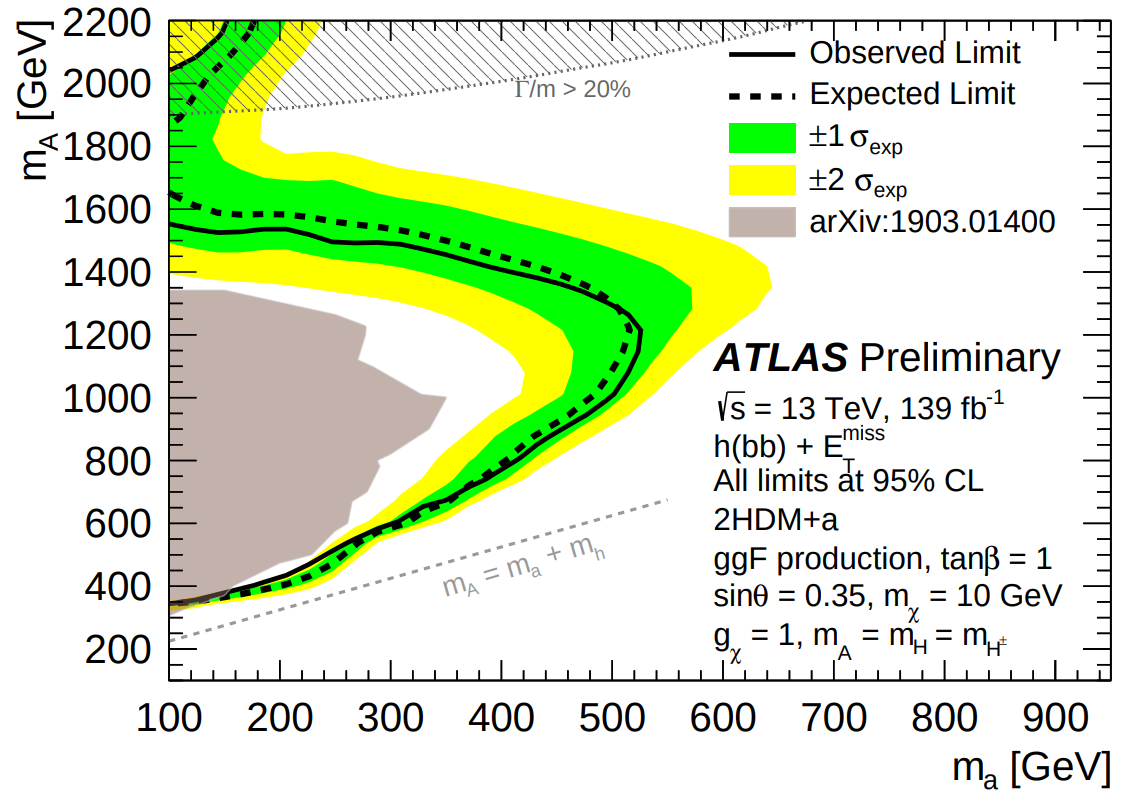}
	\caption{95\% CL exclusion contours for the 2HDM+a signal~\cite{ATLASmonoH}.}
	\label{fig:2HDMaContour}
\end{figure}

The data show no significant deviation from the SM prediction.
Figure~\ref{fig:2HDMaContour} shows the 95\% confidence level exclusion contour for the 2HDM+a model~\cite{2HDMapaper}.

\FloatBarrier

\section{Conclusion}

A wide range of analyses focusing on final states with jets to explore the Dark Sector is presented.
The data used in these analyses was collected by the ATLAS and CMS detectors at the LHC.
Both experiments have collected $\approx 140\,\mathrm{fb}^{-1}$ of proton-proton collision data at a centre-of-mass energy of 13 TeV during Run 2.

The data show no evidence for DM.
Besides preparing for the upcoming Run 3 to collect more data the two collaborations are also working on improving their analysis techniques as well as exploring new final states.

\section*{Copyright}
Copyright 2021 CERN for the benefit of the ATLAS and CMS Collaborations. Reproduction of this article or parts of it is allowed as specified in the CC-BY-4.0 license.